\documentclass[letterpaper]{article} % DO NOT CHANGE THIS
\usepackage{aaai25}  % DO NOT CHANGE THIS
\usepackage{times}  % DO NOT CHANGE THIS
\usepackage{helvet}  % DO NOT CHANGE THIS
\usepackage{courier}  % DO NOT CHANGE THIS
\usepackage[hyphens]{url}  % DO NOT CHANGE THIS
\usepackage{graphicx} % DO NOT CHANGE THIS
\usepackage{natbib}  % DO NOT CHANGE THIS AND DO NOT ADD ANY OPTIONS TO IT
\usepackage{caption} % DO NOT CHANGE THIS AND DO NOT ADD ANY OPTIONS TO IT
\usepackage{algorithm}
\usepackage{algorithmic}

\usepackage{newfloat}
\usepackage{url}            % simple URL typesetting
\usepackage{booktabs}       % professional-quality tables
\usepackage{amsfonts}       % blackboard math symbols
\usepackage{nicefrac}       % compact symbols for 1/2, etc.
\usepackage{microtype}      % microtypography
\usepackage{multirow}
\usepackage{graphicx}
\usepackage{xspace}
\usepackage{amsmath}
\usepackage{amssymb}
\usepackage{tikz}
\usepackage{pifont}
\usepackage{newfloat}
\usepackage{listings}
\newcommand{\ie}{{\emph{i.e.}},\xspace}

\newcommand{\eg}{{\emph{e.g.}},\xspace}

\newcommand{\cmark}{\ding{51}}%
\newcommand{\xmark}{\ding{55}}%
\newcommand{\greencheck}{{\cmark}}
\newcommand{\redcross}{{\xmark}}
\newcommand{\vModelName}{\textit{Draw an Audio}}

\DeclareCaptionStyle{ruled}{labelfont=normalfont,labelsep=colon,strut=off} % DO NOT CHANGE THIS
\floatstyle{ruled}
\newfloat{listing}{tb}{lst}{}
\floatname{listing}{Listing}
\title{Draw an Audio: Leveraging Multi-Instruction for Video-to-Audio Synthesis}
\author{
    Qi Yang\textsuperscript{\rm 1,2}\thanks{Work done during an internship at Meituan.},
    Binjie Mao\textsuperscript{\rm 3},
    Zili Wang\textsuperscript{\rm 1,2},
    Xing Nie\textsuperscript{\rm 1,2},
    Pengfei Gao\textsuperscript{\rm 3}, \\
    Ying Guo\textsuperscript{\rm 3},
    Cheng Zhen\textsuperscript{\rm 3},
    Pengfei Yan\textsuperscript{\rm 3},
    Shiming Xiang\textsuperscript{\rm 1,2}
}
\affiliations{
    \textsuperscript{\rm 1}~School of Artificial Intelligence, University of Chinese Academy of Sciences~(UCAS) \quad \\ 
    \textsuperscript{\rm 2}~Institute of Automation, Chinese Academy of Sciences~(CASIA) \quad \textsuperscript{\rm 3}~Meituan

}

\usepackage{bibentry}
\begin{document}

\maketitle

\begin{abstract}
Foley is a term commonly used in filmmaking, referring to the addition of daily sound effects to silent films or videos to enhance the auditory experience.
Video-to-Audio~(V2A), as a particular type of automatic foley task, presents inherent challenges related to audio-visual synchronization. 
These challenges encompass maintaining the content consistency between the input video and the generated audio, as well as the alignment of temporal and loudness properties within the video.
To address these issues, we construct a controllable video-to-audio synthesis model, termed \textit{Draw~an~Audio}, which supports multiple input instructions through drawn masks and loudness signals.
To ensure content consistency between the synthesized audio and target video, we introduce the Mask-Attention Module~(MAM), which employs masked video instruction to enable the model to focus on regions of interest.
Additionally, we implement the Time-Loudness Module~(TLM), which uses an auxiliary loudness signal to ensure the synthesis of sound that aligns with the video in both loudness and temporal dimensions.
Furthermore, we have extended a large-scale V2A dataset, named VGGSound-Caption, by annotating caption prompts.
Extensive experiments on challenging benchmarks across two large-scale V2A datasets verify \textit{Draw an Audio} achieves the state-of-the-art. Project page:~\url{https://yannqi.github.io/Draw-an-Audio/}. 
\end{abstract}

\section{Introduction}
Foley design, a prevalent profession in film-making since the 1920s, involves adding specific sound effects to silent films, videos, or other media to provide a better experience for the audience.
Generally, traditional foley sounds are always meticulously crafted by individuals mimicking real sound sources, \eg \textit{thunder can be imitated via bending thin metallic plates} and \textit{the sound of footsteps in snow can be generated by squeezing cornstarch in leather}.
As a result, producing sounds for numerous silent videos is time-intensive and expertise-required.
Recently, with the success of video generation~\cite{blattmann2023stable, Brooks2024Sora}, designers can easily generate video clips according to their demands.
However, most of the synthesis clips are soundless, diminishing the appeal of the videos. 
To bridge the gap between real videos and generated videos in terms of sound, automatic video foley technology has gradually attracted significant interest in both academia and industry~\cite{zhou2018visual,chen2020generating,liu2023sounding,luo2024difffoley}.

\begin{figure*}[ht]
\centering
\vspace{-0.3cm}
\includegraphics[scale=0.36]{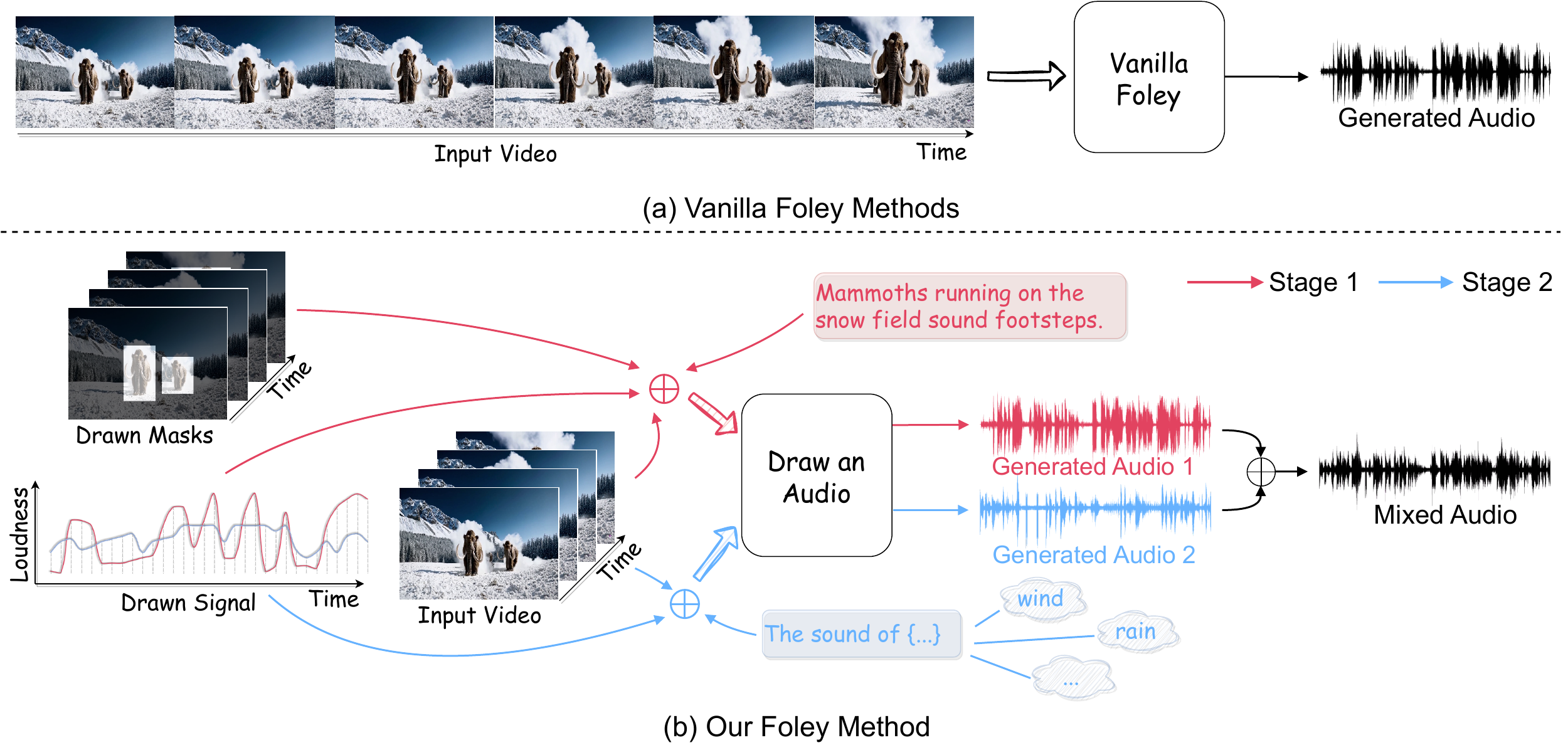}  
\caption{Schematic illustrations of vanilla foley methods and our method. The traditional methods produce the entire audio only from video inputs once, demonstrating limitations in controllability and flexibility.
\textit{Draw an Audio}, in contrast, offers a more appealing alternative that employs multiple instructions to produce high-quality synchronized audio and can produce mixed audio in multi-stages, thereby exhibiting greater practical application.
}
\vspace{-0.2cm}
\label{fig:figure0}
\end{figure*}

Fortunately, recent advances in Latent Diffusion Model~(LDM)~\cite{2022high-resolution} have brought to many promising models for content synthesis, such as Text-to-Image~(T2I) generation~\cite{2022hierarchical, 2022high-resolution}, Text-to-Audio~(T2A) generation ~\cite{ghosal2023text, liu2023audioldm, liu2023audioldm2, evans2024fast} and Video-to-Audio~(V2A) generation~\cite{luo2024difffoley,xing2024seeing, wang2024v2a}.
Despite previous methods being impressive, as an audio generation task involving video, V2A generation still presents several challenges:
(1)~Content consistency. 
It is significant for foley designers to ensure the generated sounds align with the semantic content of the corresponding videos, \ie dog barking should not appear in a cat-only video;
(2)~Temporal consistency. 
Different from the T2A generation that only concentrates on the consistency of audio content with the text input, V2A necessitates the produced foley sound to be temporally synchronized with the video;  
(3)~Loudness consistency. 
Considering human auditory sensitivity to loudness variations~\cite{fletcher1933loudness}, it is essential to generate sounds that match the appropriate intensity levels involving video. For instance, the sound of the elephant's footsteps should gradually increase as the elephant approaches in the video.

To address the above challenges in the V2A generation task, recent studies~\cite{iashin2021taming,sheffer2023hear,luo2024difffoley,wang2024v2a} attempt to generate sound aligned with video content by directly utilizing visual features for issue (1).
Though impressive, these methods are only able to make entire audio from a single video once, lacking the controllability of temporal and loudness.
Addressing issue (2), SonicVisionLM~\cite{xie2024sonicvisionlm} takes the timestamp condition into account for better temporal consistency to a certain extent while neglecting the visual perception since the input only contains the text prompt.
Issue~(3) draws less attention and still demands a full exploration.

In this work, we propose a more controllable audio synthesis framework, termed \textit{Draw an Audio}, which effectively addresses the content, temporal, and loudness consistency simultaneously.
As shown in Fig.~\ref{fig:figure0}, \textit{Draw an Audio} is capable of generating synthesized audio with multiple instructions, such as video, text, drawn video masks, and drawn loudness signals.
Drawing inspiration from Animate-Anything~\cite{dai2023animateanything}, we design Mask-Attention Module~(MAM) to enhance the semantic content consistency between the generated audio and input video.
By adaptively incorporating drawn video masks into the original video, \textit{Draw an Audio} concentrates on the regions of interest with more controllability.
Furthermore, to maintain consistency in both temporal and loudness during video foley, we introduce the Time-Loudness Module~(TLM). 
Specifically, by supporting a hand-drawn loudness signal as input, TLM endows \textit{Draw an Audio} with generating audio with specific loudness variations.
To facilitate the training of TLM, an enhanced algorithm integrating both Root Mean Square~(RMS)~\cite{panagiotakis2005speech} and Exponentially Weighted Moving Average~(EWMA)~\cite{lucas1990exponentially} is designed to convert the original audio into a handcraft-like signal.
Besides, for the training of \textit{Draw an Audio}, we further annotate VGGSound~\cite{chen2020vggsound} with caption labels and extend a new dataset named VGGSound-Caption.
During the inference, \textit{Draw an Audio} is able to synthesize mixed audio in multi-stages, exhibiting greater practical application. More visualization is in the project page.

The key contributions are highlighted as follows:
\begin{itemize}
\item We propose \textit{Draw an Audio} that supports multiple instructions as inputs. The proposed method generates audios with superior content, temporal and loudness consistency, catering to the demand of foley designers more effectively.
\item Through the implementation of the Mask-Attention Module and Time-Loudness Module, additional masks and loudness signals are introduced as instructions for designers to make the sound intensity more controllable.
\item Comprehensive experiments conducted on our proposed dataset, as well as other challenging datasets, adequately demonstrate that \textit{Draw an Audio} achieves superior performance and broader application.
\end{itemize}

\section{Related Work}

\subsection{Latent Diffusion Models}
Diffusion models~\cite{ho2020denoising,song2020denoisingddim} have demonstrated exceptional generative capabilities through the denoising processes.
Recently, the Latent Diffusion Model~(LDM)~\cite{2022high-resolution} offers an efficient and effective method to enhance both the training and sampling efficiency significantly. 
By mapping the denoising diffusion features into a latent space, LDM considerably reduces the computation without compromising the quality.
Furthermore, by leveraging a cross-attention conditioning mechanism, LDM supports diverse generative tasks, including image generation~\cite{2022high-resolution, chen2023pixartalpha, sauer2024fast, Feng_2023_CVPR}, video generation~\cite{blattmann2023stable, xing2023dynamicrafter, dai2023animateanything, guo2023animatediff, blattmann2023stable, Brooks2024Sora} and audio generation~\cite{liu2023audioldm, liu2023audioldm2, evans2024fast, ghosal2023text, huang2023make, luo2024difffoley}.
These advancements in latent diffusion-based techniques significantly inspire our research on Video-to-audio generation.

\begin{figure*}[t]
\centering
\vspace{-0.2cm}
\includegraphics[scale=0.5]{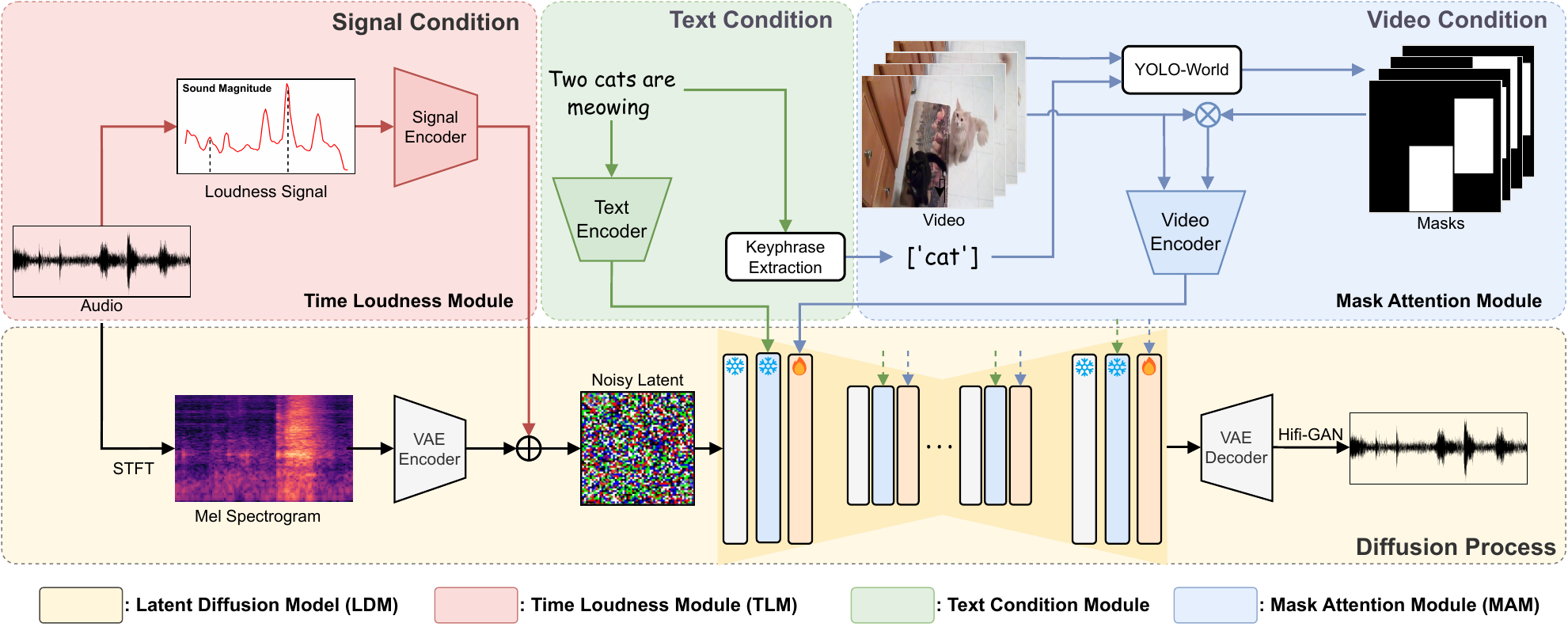} 
\caption{The architecture of \textit{Draw an Audio}, which incorporates a Latent Diffusion Model (LDM) as the foundational model, a Text Condition Model for text instruction, a Masked-Attention Module~(MAM) for video instruction, a Time-Loudness Module~(TLM) for signal instruction. Best viewed in color.}
\vspace{-0.2cm}
\label{fig:figure1}
\end{figure*}

\subsection{Text-to-Audio Generation}
Text-to-Audio generation~\cite{yang2023diffsound, huang2023make,liu2023audioldm,liu2023audioldm2,evans2024fast,ghosal2023text} recently gains significant attention.
DiffSound~\cite{yang2023diffsound} employs a discrete diffusion model to generate discrete tokens, serving as a compressed representation of mel-spectrograms with less complexity and inference speed.
Make-An-Audio ~\cite{huang2023make} enhances audio production by employing a distill-then-reprogram approach with varied concept composition and utilizes CLAP~\cite{elizalde2023clap} for deep natural language understanding.
TANGO~\cite{ghosal2023text} harnesses latent diffusion models, and AudioLDM2~\cite{liu2023audioldm2} further incorporates AudioMAE~\cite{huang2022masked} to improve the quality of the generated audio.
In this study, we take AudioLDM2~\cite{liu2023audioldm2} as our latent diffusion model architecture for further exploration.

\subsection{Video-to-Audio Generation}
The field of Video-to-Audio generation~\cite{owens2016visually,zhou2018visual,chen2020generating,iashin2021taming,liu2023sounding,luo2024difffoley,xie2024sonicvisionlm} is exhibiting significant potential, paralleling advances in the area of Text-to-Video generation~\cite{Brooks2024Sora,blattmann2023stable}.
Im2Wav~\cite{sheffer2023hear} incorporates a transformer model and uses CLIP~\cite{radford2021learning} features as a condition but suffers from slow inference speed. 
SpecVQGAN~\cite{iashin2021taming} deploys a transformer to produce new spectrograms from a pre-trained codebook based on the input video features.
The Diff-Foley model~\cite{luo2024difffoley} employs a latent diffusion model to generate high-quality audio and utilizes contrastive audio-visual pre-training to learn features to achieve better temporal and semantic alignment.
In contrast, SonicVisionLM~\cite{xie2024sonicvisionlm} leverages the capabilities of vision language models to suggest sounds that correspond to the video content. 
However, current V2A methods encounter challenges with synchronization because they rely solely on video or text conditions, limiting the capture of intricate audio-visual correlations.

\section{Method}

\textit{Draw an Audio} aims to generate sounds that align with the input videos in terms of content, timing, and loudness.
As illustrated in Fig.~\ref{fig:figure1}, the model of \textit{Draw an Audio} is denoted as $\mathcal{F} : \boldsymbol{\tau} \rightarrow \boldsymbol{x}_a$, wherein $\mathcal{F}$ represents the entire function, $\boldsymbol{x}_a$ is the generated audio and $\boldsymbol{\tau}$ denotes the involved conditioning sets. 
To understand our model comprehensively, we first outline the basic framework of \textit{Draw an Audio}.
Subsequently, we detaile the components of Mask-Attention Module~(MAM) and Time-Loudness Module~(TLM), respectively.

\subsection{Latent Diffusion Model}

Considering that Latent Diffusion Model~(LDM)~\cite{2022high-resolution} has achieved significant success in Text-to-Audio~(T2A)~\cite{liu2023audioldm, ghosal2023text} and Video-to-Audio~(V2A)~\cite{luo2024difffoley, wang2024v2a} fields, we select the same latent diffusion architecture as the foundation model due to its excellent performance.

Following AudioLDM2~\cite{liu2023audioldm2}, we employ log mel-spectrograms $\boldsymbol{x}_{spec} \in \mathbb{R}^{T' \times M}$, transformed from the audio $\boldsymbol{x}_a$, as the input. 
Here, $M$ denotes the mel-bins, and $T'$ represents the time axis. 
Notably, the LDM learns the reverse diffusion process in a compressed latent space, $\boldsymbol{z} \in \mathbb{R}^{h \times w \times c}$, derived from $\boldsymbol{x}_{spec}$ through a variational autoencoder~(VAE)~\cite{kingma2013auto}, effectively reducing the computational burden.
Formally, a latent feature $\boldsymbol{z}_0$ is initially sampled from a real data distribution $\boldsymbol{z}_0 \sim q(\boldsymbol{z})$.
Subsequently, LDM performs $T$-step forward diffusion process with a Markov chain to slowly add Gaussian noise with scheduled noise parameters $0 < \beta_1 < \beta_2 < \ldots < \beta_T < 1$. 
The forward diffusion process can be formulated as follows:
\begin{equation}
q(\boldsymbol{z}_t|\boldsymbol{z}_{t-1}) = \sqrt{1-\beta_t} \boldsymbol{z}_{t-1} + \sqrt{\beta_t} \boldsymbol{\epsilon}_t, 
\label{eq:forward_diffusion_process_1}
\end{equation}
\begin{equation}
q(\boldsymbol{z}_t|\boldsymbol{z}_{0}) = \sqrt{\alpha_t} \boldsymbol{z}_{0} + \sqrt{1-\alpha_t} \boldsymbol{\epsilon}_t, 
\label{eq:forward_diffusion_process_2}
\end{equation}
where $\boldsymbol{\epsilon} \sim \mathcal{N}(\mathbf{0}, \boldsymbol{I})$, and $\alpha_t = \prod_{t=1}^{t}(1-\beta_t)$ represents the recursive substitution of $q(\boldsymbol{z}_t|\boldsymbol{z}_{t-1})$ in Eq.~\ref{eq:forward_diffusion_process_1} by reparameterization trick~\cite{song2020denoisingddim}. The final step of the forward process yields $\boldsymbol{z}_T \sim \mathcal{N}(\mathbf{0}, \boldsymbol{I})$.

For the reverse diffusion process during training, LDM denoises and reconstructs $\boldsymbol{z}_0$ from the noise estimation $\boldsymbol{\hat{\epsilon}}_\theta$, which can be calculated with the following loss function:
\begin{equation}
    \mathcal{L}_{LDM} = \sum_{t=1}^T \mathbb{E}_{\boldsymbol{\epsilon_t} \sim \mathcal{N}(\mathbf{0}, \boldsymbol{I}), \boldsymbol{z}_0} || \boldsymbol{\epsilon}_t - \boldsymbol{\hat{\epsilon}}_{\theta}^{(t)}(\boldsymbol{z}_t, \boldsymbol{\tau}) ||^2_2,
    \label{eq:LDM_loss}
\end{equation}
where $\boldsymbol{z}_t$ is the prior distribution sampled from Eq.~\ref{eq:forward_diffusion_process_2} with Gaussian noise $\boldsymbol{\epsilon_t}$. $\boldsymbol{\tau}$ is the condition sets for guidance, and $\boldsymbol{\hat{\epsilon}}_{\theta}^{(t)}$ is predicted by the UNet-like architecture~\cite{ronneberger2015unet}.

During inference, LDM generates the initial audio latent $\boldsymbol{z}_0$ by sampling through the reverse process. 
This process begins with $\boldsymbol{z}_T$, which is sampled from a standard normal noise distribution, and utilizes the conditional embedding sets $\boldsymbol{\tau}$, as detailed in the following steps:
\begin{equation}
    \label{eq:ldm_reverse1}
    p_\theta(\boldsymbol{z}_{0:T}|\boldsymbol{\tau}) = p(\boldsymbol{z}_T) \prod_{t=1}^T p_\theta(\boldsymbol{z}_{t-1}| \boldsymbol{z}_t, \boldsymbol{\tau}),
\end{equation}
\begin{equation}
    \label{eq:ldm_reverse2}
    p_\theta(\boldsymbol{z}_{t-1}|\boldsymbol{z}_t, \boldsymbol{\tau}) = \mathcal{N}(\boldsymbol{z}_{t-1};\boldsymbol{\mu}_\theta(\boldsymbol{z}_t, t, \boldsymbol{\tau}), \sigma_t^2 \boldsymbol{I}),
\end{equation}
where mean and variance are parameterized~\cite{ho2022imagen} as $\boldsymbol{\mu}_\theta(\boldsymbol{z}_t, t, \boldsymbol{\tau})= \frac{1}{\sqrt{\alpha_t}}(\boldsymbol{z_t}-\frac{\beta_t}{\sqrt{1-\bar{\alpha}_t}}\boldsymbol{\epsilon}_\theta(\boldsymbol{z}_t,t,\boldsymbol{\tau}))$ and $\sigma^2_t = \frac{1-\bar{\alpha}_{t-1}}{1-\bar{\alpha}_t}\beta_t$ with $\sigma_1^{2} = \beta_1$.
Among the condition sets $\boldsymbol{\tau}$, we provide the multiple instructions in \textit{Draw an Audio}, \ie $\boldsymbol{\tau} \in \{\boldsymbol{\tau}_{text}, \boldsymbol{\tau}_{visual}, \boldsymbol{\tau}_{signal}\}$, which present the text condition, video condition, and loudness signal condition, respectively.

% AudioCaps
\begin{table*}[ht!]
\centering
% \small
\begin{tabular}{l|c|cc|ccccc}
\toprule
Model & Train Data. & Video & Text &  FD$\downarrow$ & IS$\uparrow$ & KL$\downarrow$  & FAD$\downarrow$ & CS-AV$\uparrow$  \\
\midrule
SpecVQGAN~\cite{iashin2021taming} & 1,3 & \checkmark &  & 44.5 & 4.52 & 3.50 & 5.34 & 6.60\\
AudioLDM2~\cite{liu2023audioldm2} &1,2,4,5 & & \checkmark & 51.8 & 6.54 & 1.67 & \textbf{1.96}  & 8.58  \\
Im2Wav~\cite{sheffer2023hear}  & 1 & \checkmark &  & 35.3 & 5.76 & 2.58 & 6.87 & 8.54 \\
Seeing and Hearing~\cite{xing2024seeing} & - & \checkmark & \checkmark & 44.6 & 4.40 & 2.80 & 8.13 & 6.11 \\
\textsc{DIFF-FOLEY}~\cite{luo2024difffoley}  & 1,4  & \checkmark &  & 40.8 & 9.14 & 3.28 & 7.15 & 7.85 \\

\textbf{\vModelName} & 2 & \checkmark & \checkmark & 24.2 & 9.33 & 1.31 & 3.26 & 8.92 \\
\textbf{\vModelName\textit{-Full}}  & 1*,2 & \checkmark & \checkmark & \textbf{23.7} & \textbf{9.52} & \textbf{1.30} & 3.89 & \textbf{9.24} \\
\midrule
\end{tabular}
\caption{Comparison of model performances on the AudioCaps evaluation set.  Models marked with \textit{Full} are trained solely on the AudioCaps and VGGSound-Caption datasets.  `Train Data.` column lists datasets as follows: 1: VGGSound~\cite{chen2020vggsound}, where 1* denotes our VGGSound-Caption  2: AudioCaps~\cite{kim2019audiocaps}, 3: VAS Dataset~\cite{chen2020generating}, 4: AudioSet~\cite{gemmeke2017audio}, 5: WavCaps~\cite{mei2024wavcaps}.}

\vspace{-0.2cm}
\label{tab:audiocaps_main}
\end{table*}

\subsection{Mask-Attention Module~(MAM)}
\label{sec:MAM}
Considering vanilla foley methods~\cite{iashin2021taming,luo2024difffoley,wang2024v2a,xie2024sonicvisionlm} generate audio directly without addressing the local information in the video, it is difficult to produce specific sounds that align with targeted content, \eg selectively focusing on the sounds of cats in a video.
To address this issue, we design the Mask-Attention Module~(MAM) with enhanced area controllability.
Specifically, by leveraging the prior masks of the video, our model enhances the region of interest, thereby ensuring content consistency.

Before implementing the MAM, it is essential to consider acquiring appropriate masks readily.
Given the text-video pair $(\boldsymbol{x}_t,\boldsymbol{x}_v)$, where $\boldsymbol{x}_t$ represents the caption prompt, and  $\boldsymbol{x}_v \in \mathbb{R}^{T_f \times H \times W \times 3}$ denotes the input video clip with $T_f$ frames, our goal is to obtain a substantial number of binary masks corresponding to sound-related regions in videos.
To achieve this, we utilize YOLO-World~\cite{cheng2024yolo} to generate the sound-related masks using an open vocabulary in real time.
Specifically, given that the caption $\boldsymbol{x}_t$ is inappropriate for the open-vocabulary model, we initially adopt the simple n-gram algorithm to extract the noun phrases list. Subsequently, these extracted noun phrases are fed into a text encoder pre-trained by CLIP~\cite{radford2021learning} to extract the corresponding text embeddings $\boldsymbol{X}_t \in \mathbb{R}^{C \times D}$, where $C$ is the number of nouns, and $D$ denotes the embedding dimension.
Formally, we can obtain the binary masks by:
\begin{equation}
    \boldsymbol{M} = \mathcal{F}_{\textit{YOLO-World}}(\boldsymbol{x}_v, \boldsymbol{X}_t),
    \label{eq:yolomask}
\end{equation}
where $ \mathcal{F}_{\textit{YOLO-World}}$ represents the YOLO Detector, mainly based on YOLOv8~\cite{jocheryolov8}, and $\boldsymbol{M} \in \mathbb{R}^{T_f  \times H \times W \times 1}$ denotes the binary masks related to the caption prompt.

After extracting the masks from the video, we further expect to inject the masked video features into MAM while preserving the entirety of the video information. 
Besides, given the cross-modal nature of our task, which involves both video and audio, we employ CAVP~\cite{luo2024difffoley} as visual encoder $E_v$.
Unlike traditional methods~\cite{sheffer2023hear,xing2024seeing}, CAVP is analogous to CLIP but has been trained specifically on audio-video dataset pairs.
To effectively integrate the masked features into \textit{Draw an Audio} while retaining the original video features, we introduce channel-weighted blocks that augment the original visual features to obtain the visual condition encoding $\boldsymbol{\tau}_{visual} \in \mathbb{R}^{ T_f \times C}$. 
The formula can be written as follows:
\begin{equation}
    \boldsymbol{F}_{visual} = E_v(\boldsymbol{x}_v); \quad    \boldsymbol{F}_{mask} = E_v(\boldsymbol{M} \odot \boldsymbol{x}_v),
    \label{eq:cavpmask1}
\end{equation}
\begin{equation}
    \boldsymbol{\tau}_{visual} = \boldsymbol{F}_{mask}(\text{GAP}(\boldsymbol{F}_{mask})\boldsymbol{W}) + \boldsymbol{F}_{visual}, 
    \label{eq:cavpmask2}
\end{equation}
where $\boldsymbol{F}_{visual}$ and $\boldsymbol{F}_{mask} \in \mathbb{R}^{T_f \times C}$ are the original video and masked video features, respectively. 
$\text{GAP}(\cdot)$ stands for global average pooling, $\boldsymbol{W} \in \mathbb{R}^{C \times C} $ represent the linear weight. For simplicity, the bias is omitted in this context. 

Having obtained the video condition encoding $\boldsymbol{\tau}_{visual}$ with masked attention to boost content consistency, we further insert it into \textit{Draw an Audio} with cross-attention to enhance the UNet features, which can be defined as:
\begin{equation}
\left \{
\begin{array}{ll}
    \mathbf{Q} = \mathbf{W}_{Q} \cdot \boldsymbol{z}';\ \mathbf{K} = \mathbf{W}_{K}\cdot \boldsymbol{\tau}_{visual};\ \mathbf{V} = \mathbf{W}_{V}\cdot \boldsymbol{\tau}_{visual}; \\
    \text{CrossAttention}(\mathbf{Q}, \mathbf{K}, \mathbf{V}) = \text{softmax}(\frac{\mathbf{Q}\mathbf{K}^T}{\sqrt{d}})\cdot \mathbf{V},
\end{array}
\right.
\label{eq:ldm_ca}
\end{equation}
where $ \boldsymbol{z}' \in \mathbb{R}^{h' \times w' \times c'}$ represents the middle latent features, $h'$ the height, $w'$ the width and $c'$ the number of channels. And $\mathbf{W}_{Q}$, $\mathbf{W}_{K}$, $\mathbf{W}_{V}$ are the linear weight for mapping into the same dimension.
% Note that text embeddings are duplicated for video frames.

\subsection{Time-Loudness Module~(TLM)}

The essence of video foley is to generate the audio with greater time alignment and loudness consistency to meet the demands of foley designers.
To pursue this goal, we believe that introducing hand-drawn loudness signals as input conditions is an excellent way to maintain temporal and loudness consistency by indirectly controlling.
However, the challenge in training \textit{Draw an Audio} with such signal conditions lies in the scarcity of large datasets containing smooth signals aligned with video-audio pairs.

Consequently, we innovatively propose an audio transformation approach based on root mean square~(RMS)~\cite{panagiotakis2005speech}. 
This approach converts the original audio into human-like, hand-drawn loudness signals that serve as part of the input instruction to our model.
The RMS energy, which considers the energy of the waveform, provides a more accurate representation of the average loudness of the audio compared to peak levels.

Specifically, define $\boldsymbol{x}_a \in \mathbb{R}^{L_a}$ as the audio signal, where $L_a$ represents the audio length. The formula for RMS energy of audio signal $\boldsymbol{F}_{{rms}}$, is defined as follows:
\begin{equation}
    \begin{aligned}
    \boldsymbol{F}_{{rms}}(i) & = \sqrt{\frac{1}{N_{{win}}} \sum_{n=i}^{i+N_{{win}} -1} (\boldsymbol{x}_a(n))^2}, \\
    \text{for } i & = 0, N_{{hop}}, 2N_{{hop}}, ..., L_a - N_{{win}},
    \end{aligned}
\end{equation}
where $N_{{win}}$ represents the \textit{window size} indicating the number of samples in each frame.  
$N_{{hop}}$ denotes the \textit{hop length}, which refers to the number of sample shifts that occur when transitioning from one frame to the next.
And $\boldsymbol{F}_{{rms}} \in \mathbb{R}^{L_{rms}}$, where $L_{rms} = \frac{L_a - N_{win}}{ N_{hop}} + 1$. Note that zero-padding is employed for the signals when $\frac{L_a - N_{win}}{ N_{hop}}$ is not divisible.

To emulate a hand-drawn signal more closely, we perform adaptive average pooling~(AAP) and exponentially weighted moving average~(EWMA)~\cite{lucas1990exponentially}, further simplifying and smoothing the signal.
Initially, we compress the signal into a lower-dimensional form using  $\boldsymbol{F}'_{rms} = \text{AAP}(\boldsymbol{F}_{rms}) \in \mathbb{R}^{L'_{rms}}$, where $\mathbb{R}^{L'_{rms}} \ll \mathbb{R}^{L_{rms}}$, to facilitate easier manual plotting.
Then, the EWMA is applied to obtain a signal that closely resembles a hand-drawn one, as follow:
\begin{equation}
\boldsymbol{F}_{signal} = \sum_{i=-N'_{win}/2}^{N'_{win}/2} \boldsymbol{F}'_{rms}(t-i) \cdot w'(i),   
\end{equation}
where $w'(\cdot)$ is the Gaussian kernel after normalization with a variance of $\sigma$, $N'_{win}$ is the \textit{window size}. 
After obtaining the loudness signal, we next use the MLP layer as the signal encoder to map $\boldsymbol{F}_{signal}$ as the $\boldsymbol{\tau}_{signal} \in \mathbb{R}^{1 \times w \times 1}$ and then add it into $\boldsymbol{z}$ for training, where $\boldsymbol{z} = \boldsymbol{z} + \boldsymbol{\tau}_{signal}$. 

During the inference phase, foley designers can flexibly control the loudness of the generated audio by manually drawing loudness signals. 
TLM enhances consistency in terms of both temporal alignment and loudness. Further details and visualizations of the loudness signal production process are available in our Supplementary Material.

\subsection{Text and Video Classifier-Free Guidance}
During the inference phase, \textit{Draw an Audio} makes sounds relying on text, video, and loudness signal simultaneously.
For the text and video conditions, we employ the dual classifier-free guidance strategy to modulate the strength of both two conditions.
Classifier-free guidance~\cite{ho2022classifier} is a mechanism in sampling that pushes the distribution of predicted noise towards the conditional distribution using an implicit classifier.
Specifically, the conditional prediction is used to guide the unconditional prediction towards the conditional, which is as follow:
\begin{equation}
    \begin{aligned}
    \boldsymbol{\hat{\epsilon}}_{\theta}^{(t)}(\boldsymbol{z}_t) &=  \boldsymbol{\hat{\epsilon}}_{\theta}^{(t)}(\boldsymbol{z}_t, \varnothing,   \varnothing) \\
    & + s_{text} \cdot (\boldsymbol{\hat{\epsilon}}_{\theta}^{(t)}(\boldsymbol{z}_t, \boldsymbol{\tau}_{text}, \varnothing) - \boldsymbol{\hat{\epsilon}}_{\theta}^{(t)}(\boldsymbol{z}_t, \varnothing,   \varnothing)) \\
    & + s_{video} \cdot (\boldsymbol{\hat{\epsilon}}_{\theta}^{(t)}(\boldsymbol{z}_t, \boldsymbol{\tau}_{text}, \boldsymbol{\tau}_{video}) - \boldsymbol{\hat{\epsilon}}_{\theta}^{(t)}(\boldsymbol{z}_t, \varnothing,  \varnothing)),
    \end{aligned}
\end{equation}
where $s_{text}$ and $s_{video}$ denote the classifier-free-guidance scalar weights. A more comprehensive discussion about the scalar weights is detailed in the ablation experiment.

\section{Experiments}

\begin{figure*}[ht!]
\centering
% \vspace{-0.1cm}
\includegraphics[scale=0.5]{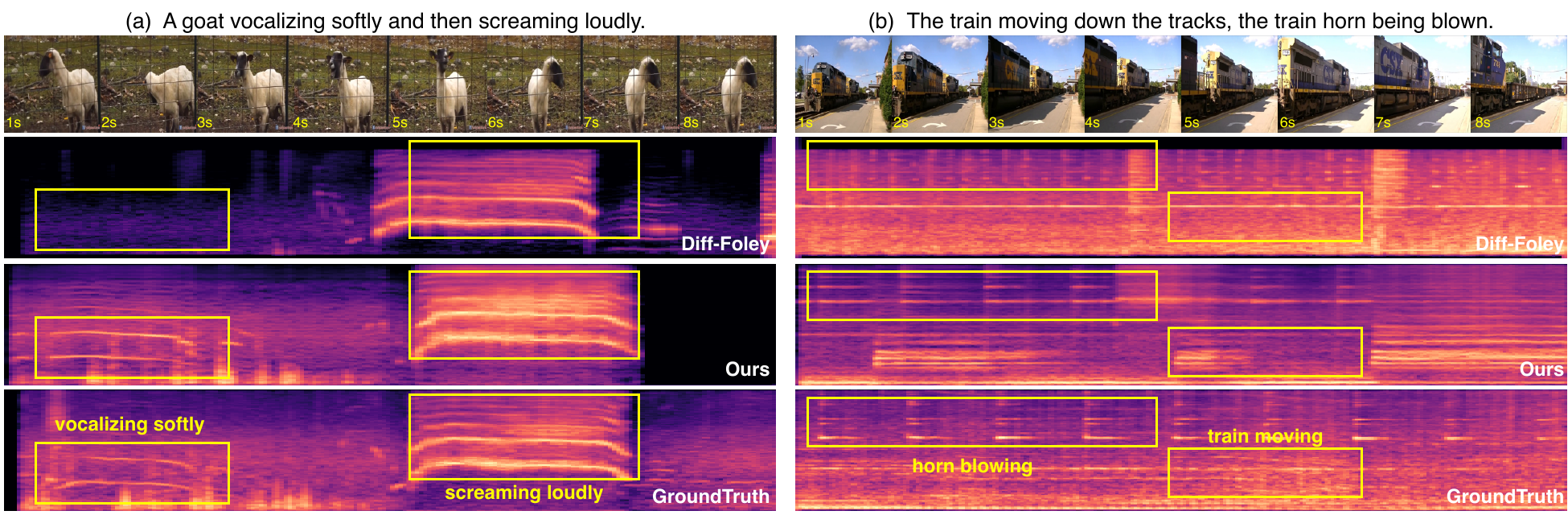} 
\caption{ 
Visualization of generated samples. 
In fig.~(a), it can be seen that while the other method fails to generate the temporal synchronized audio based on the video, \textit{Draw an Audio} successfully produces both quiet and loud sound with temporal consistency. In fig.~(b), our method can generate the sound with clear spikes matching the ground truth mel-spectrogram.} 
\label{fig:figure_visualization}
\vspace{-0.1cm}
\end{figure*}

\subsection{Experimental Setup}
\paragraph{Datasets.} 
Our experiments are conducted using the VGGSound-Caption and AudioCaps~\cite{kim2019audiocaps} datasets. 
The VGGSound-Caption dataset, as an expanded version of the VGGSound~\cite{chen2020vggsound} dataset, comprises approximately 200K 10-second video clips and additional caption prompts across more than 300 classes. 
The AudioCaps dataset contains around 46K 10-second video clips with natural language descriptions in the wild through crowdsourcing on the AudioSet~\cite{gemmeke2017audio} dataset.
Actually, we downloaded a total of 223k 10-second video clips from the dataset for the experiments, as some of the videos were removed from YouTube.
During the evaluation, we randomly sampled 1000 video clips from the test sets of the two datasets to verify the performance respectively.
Our model is trained and evaluated separately on these two datasets. 
To enhance practical application, we also conduct training on a mixed dataset incorporating both sources.

\begin{table}[tbp]
\centering
\small

\begin{tabular}{l|ccccc}
\toprule
Model &  FD$\downarrow$ & IS$\uparrow$ & KL$\downarrow$  & FAD$\downarrow$ & CS-AV$\uparrow$  \\
\midrule
SpecVQGAN  & 36.9  & 4.73 & 3.43 & 5.01 & 6.76\\
AudioLDM2 & 29.2 & 8.56 & 3.14 & \textbf{2.88}  & 7.54  \\
Im2Wav & 27.4 & 5.96 & \textbf{2.42} & 6.09 & 8.86 \\
Seeing and Hearing  & 35.5 & 4.71 & 2.86 & 6.13 & 6.08 \\
\textsc{DIFF-FOLEY}  & 31.7 & 8.89 & 3.26 & 6.59 & 7.63 \\
\textbf{\vModelName} & 27.5 & 9.61 & 2.60 & 3.24 & 9.73 \\
\textbf{\vModelName\textit{-Full}}  & \textbf{25.3} & \textbf{9.98} & 2.46 & 3.33 & \textbf{9.80} \\
\midrule
\end{tabular}
\caption{Comparison of model performances on the VGGSound-Caption evaluation set. 
Models marked with \textit{Full} are exclusively trained on both two datasets.}
\label{tab:vggsound_main}
\vspace{-0.35cm}
\end{table}

\paragraph{Data Pre-Processing.}
For the video components, we employ the same processing methods as used in Diff-Foley~\cite{luo2024difffoley}.
Specifically, initial sampling of the 10-second input video clips is conducted at a rate of 4 frames per second, yielding a total of 40 frames. These frames are subsequently resized to dimensions of $224 \times 224$.
Regarding the audio components, all audio data are resampled to a frequency of 16kHz for consistency with previous studies~\cite{iashin2021taming,luo2024difffoley}.
Throughout the training phase, we randomly crop video-audio clips to a length of 8 seconds, which results in $T_f = 32$ and $L_a = 128000$.

\paragraph{Hyper-parameters and Training Details.} 
The latent diffusion model of \textit{Draw an Audio} is constructed on the architecture of AudioLDM2~\cite{liu2023audioldm2}.
Therefore, we employ the same training setting following~\cite{liu2023audioldm2}.
Specifically, the audio clips are transposed into mel-spectrograms with a mel length of $T'=800$ and mel-bins of $M=64$.
For the input of text condition, we combine the text embeddings from the frozen CLAP~\cite{wu2023large} and the frozen FLAN-T5~\cite{chung2024scaling} as our text encoder.
With regard to the video condition input, the frozen CAVP~\cite{luo2024difffoley} as the visual encoder is employed to extract the video features with the same dimension $C = 768$ as~\cite{luo2024difffoley} for fair comparison. 
Concerning the introduced loudness signal condition, we set $N_{hop}=160$ and $N_{win}=1024$ to compute the RMS energy.
Besides, we set $N'_{win}=3$, $\sigma = 3.5$ for the EWMA module map RMS energy into a lower dimension with 10 frames per second, resulting in $L'_{rms}=80$. The signal encoder consists of trainable MLP layers.
During the inference phase, the scale of the Classifier-Free Guidance is set to $s_{text}=3.4$ and $s_{video}=4.5$, respectively. The DPM-Solver~\cite{lu2022dpm} sampler is used with 25 sampling steps.
More training details are in Supplementary Material.

\paragraph{Evaluation Metrics.}
We mainly focus on the V2A generation task to evaluate the effectiveness of our \textit{Draw an Audio}.
We follow the same evaluation protocol of SpecVQGAN~\cite{iashin2021taming} and Im2Wav~\cite{sheffer2023hear}, which calculates objective metrics, such as Frechet Distance~(FD), Inception Score~(IS),  Kullback-Leibler Divergence~(KL),  Frechet Audio Distance~(FAD) and CLIP-Score of Audio and Video~(CS-AV).
Similar to the frechet inception distance in image generation, the FD in audio indicates the similarity between generated samples and target samples.
IS is effective in evaluating not only the quality of samples but also their diversity.
KL divergence measures the similarity between the generated and target audio with the label calculated by the audio tagging model, \ie Patch-out Transformer~\cite{koutini2021efficient}. 
FAD is a reference-free audio quality measure calculated based on the distribution distance between the feature of the target and generated audios, extracted from the VGGish~\cite{hershey2017cnn}.

In addition, we introduce the variations of the CLIP-SCore~\cite{hessel2021clipscore}, CS-AV, which has shown to be highly effective in evaluating audio-image correspondence.
Since our method is video audio cross-modal, we replace the text encoder of CLIP with Wav2Clip~\cite{wu2022wav2clip}, which is an audio encoder trained using contrastive loss on corresponding images and audio on the frozen CLIP image encoder.
We pass both the video with 1 frame per second and the generated sound through their respective feature extractors. Then, we compute the expectation of cosine similarity of the resultant feature vectors, multiplied by a scaling factor, $\gamma$. We set $\gamma = 100$ following Im2Wav~\cite{sheffer2023hear}.

\begin{table*}[htbp]
\small
\centering

\begin{tabular}{c|c|ccccc|ccccc}
\toprule
\multirow{2}{*}{Models} & \multirow{2}{*}{Trainable Params}  & \multicolumn{5}{c|}{AudioCaps} & \multicolumn{5}{c}{VGGSound-Caption} \\
 & & FD$\downarrow$ & IS$\uparrow$ & KL$\downarrow$  & FAD$\downarrow$ & CS-AV$\uparrow$ & FD$\downarrow$ & IS$\uparrow$ & KL$\downarrow$  & FAD$\downarrow$ & CS-AV$\uparrow$ \\
\midrule
\vModelName & 82~M &  24.2 & 9.33 & \textbf{1.31} & \textbf{3.26} & 8.92  & 27.5 & 9.61 & 2.60 & \textbf{3.24} & \textbf{9.73} \\
\vModelName* & 326~M & \textbf{23.8} & \textbf{9.74} & 1.34 & 3.42 & \textbf{8.94} & \textbf{26.2} & \textbf{11.1} & \textbf{2.53} & 3.34 & \textbf{9.73} \\

\bottomrule
\end{tabular}

\caption{Effect on the trainable parameters. \textit{Draw an Audio} only trains additional parameters based on AudioLdm2~\cite{liu2023audioldm2}, while model marked with $\ast$ are exclusively trained on the whole UNet.}
\label{tab:ablation_params}
\end{table*}

\subsection{Results}
\label{sec:results}
\paragraph{Quantitative Results.} Tab.~\ref{tab:audiocaps_main} provides quantitative results from the AudioCaps test set, including the entire training datasets of each method.
As can be seen, \textit{Draw an Audio} outperforms the baseline method significantly with fewer trainable datasets.
Specifically, our standard method achieves better objective performance in both FD~(11.3$\downarrow$), IS(0.38$\uparrow$), KL(0.37$\downarrow$) and CS-AV~(0.66$\uparrow$).
In Tab.~\ref{tab:vggsound_main}, we also verified the effectiveness of \textit{Draw an Audio} on a larger VGGSound-Caption dataset.
Besides, we train our model on both mixed two datasets, with the suffix \textit{Full}, for better application practice. 
The experimental results presented in Tab.~\ref{tab:audiocaps_main} and Tab.~\ref{tab:vggsound_main} reveal that as the duration of training data increases, the performance of \textit{Draw an Audio} can be further enhanced.

\paragraph{Qualitative Results.}
To intuitively compare the differences between \textit{Draw an Audio} and other methods, we draw the visualization from both the two datasets. 
As shown in Fig.~\ref{fig:figure_visualization}~(a), we sample a video from the AudioCaps dataset for foley. It can be seen that \textit{Draw an Audio} is able to successfully produce both quiet vocalizing and loud screaming sounds with temporal consistency. In contrast, other methods fail to create loud and temporally synchronized audio based on the video.
In~Fig.~\ref{fig:figure_visualization}~(b), we conduct our model on the VGGSound-Caption dataset. It can be seen that our method can generate the sound with clear spikes that match the ground truth mel-spectrogram, which is more precise than others.

\begin{table}[tbp!]
\vspace{-0.1cm}
\small
\centering

\begin{tabular}{ccc|ccccc}
\toprule
{Text} & {Masks} & {Signal} 
 & IS$\uparrow$&   KL$\downarrow$  &  FAD$\downarrow$  & CS-AV$\uparrow$ \\
\midrule
\redcross & \redcross & \greencheck & 6.64 & 2.62 & 5.27 & 8.62    \\
\redcross & \greencheck & \greencheck & 6.92 & 2.75 & 5.04 & 8.87  \\
\greencheck & \redcross & \redcross &  8.95 & 1.26 & 3.67 & 8.96 \\
\greencheck & \greencheck &  \redcross & 8.89 & 1.26 & 3.74 & 8.99 \\
\greencheck & \redcross & \greencheck & 8.91 & \textbf{1.25} & 3.48 & \textbf{9.01}  \\
\greencheck & \greencheck &\greencheck & \textbf{9.33}  & 1.31 & \textbf{3.26} & 8.92 \\

\bottomrule
\end{tabular}
\caption{Analysis of model performance under various input instructions. 
The validity of our proposed model is confirmed on the AudioCaps dataset by employing different combinations of text prompt, masked video, and loudness signal.
}
\label{tab:ablation_MAM}
\vspace{-0.35cm}
\end{table}

\subsection{Ablation Study}
\paragraph{Effect of Multi-Instructions.} 
\textit{Draw an Audio} allows for the selective inclusion of multiple conditions, enhancing its controllability and user-friendliness. 
As shown in Tab.~\ref{tab:ablation_MAM}, we present a comparison amongst varying input conditions such as text prompt, masked video, and loudness signal. 
Notably, text instruction has the most significant influence on \textit{Draw an Audio} as our model is pre-trained on the T2A architecture.
As the quantity of input instruction augments, the performance of our model gradually increases significantly.
This pronounced enhancement can be attributed to the integration of our proposed MAM and TLM.

\begin{figure}[tbp!]
\vspace{-0.10cm}
\centering
% \hspace{-5pt}
\includegraphics[scale=0.975]{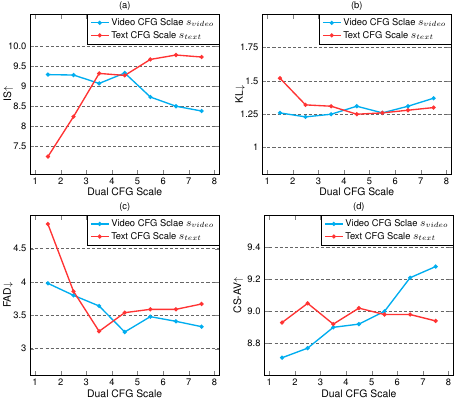} 
\caption{ 
Effect on Dual CFG Scales. Figures (a)-(d) represent various objective evaluation metrics. By maintaining $s_{text}=3.5$ and gradually increasing $s_{video}$ from $1.5$ to $7.5$, a U-shaped curve is constructed. This pattern is similarly observed when the procedure is reversed with $s_{video}=4.5$.}
\label{fig:figure_cfg}
\vspace{-0.35cm}
\end{figure}

\paragraph{Effect of Dual Classifier-Guidance Scales.} 
The dual classifier-free guidance scales are of crucial importance for sampling from latent diffusion models. 
As shown in Fig.~\ref{fig:figure_cfg}, we report the effect of varying numbers of the dual classifier-guidance~(CFG) scales $s_{text}$ and $s_{video}$ respectively.
As can be seen, the different scales have an influence on sample quality and diversity.
Considering evaluation metrics comprehensively, we set $s_{text}=3.5$ and $s_{video}=4.5$ by default.

\paragraph{Effect of Training Parameters.} 
We studied the impact of training parameters on model performance. As shown in Tab.~\ref{tab:ablation_params}, we conduct extensive experiments with more trainable parameters on each dataset.
It is evident from the results that there is a direct relationship between the increase in the number of trainable parameters and the enhancement of our model's performance.
This suggests that there is considerable scope for further development of our model.

\section{Conclusion}
We introduce \textit{Draw an Audio}, a controllable V2A approach designed to generate highly synchronized audio consistent with associated visuals.
We demonstrate our method's superiority in audio generation through comprehensive experiments.
Specifically, by integrating the Mask-Attention Module~(MAM), our model precisely directs attention to specific areas for targeted audio generation.
Through the implementation of the Time-Loudness Module (TLM), our model attains enhanced flexibility in controlling both the duration and intensity of sound.
Besides, we enhance the VGGSound dataset by annotating it with caption prompts to facilitate the training of our model, termed VGGSound-Caption.
Finally, extensive experiments and ablation analyses demonstrate the effectiveness of our proposed module and its practical application.
% \paragraph{Limitations.}
% \textit{Draw an Audio} performs great audio-visual synchronization on AudioCaps and VGGSound-Caption.
% However, due to the complexity of multi-condition inputs, our method is more suitable for professionals, \ie foley designers, rather than the common public. 

\newpage

\bibliography{aaai25}

\newpage

% \clearpage
\appendix
\setcounter{section}{0}
\setcounter{figure}{0}
\setcounter{table}{0}
\renewcommand{\thesection}{\Alph{section}} % 使section编号变为大写字母  % 用于补充材料
\renewcommand{\thefigure}{\Roman{figure}} % 使图的编号变为大写罗马数字  % 用于补充材料
\renewcommand{\thetable}{\Roman{table}} % 使表的编号变为大写罗马数字  % 用于补充材料
\section*{Appendix}

This appendix presents additional materials and visualization results.
First, we give further descriptions of our proposed Time-Loudness Module~(TLM) to enhance comprehension.
Then, we describe the detailed training settings of our model.
Finally, we provide more visualization results for \textit{Draw an Audio} for a more intuitive understanding. 
The sound and code files can be found in supplementary materials. 

\begin{figure}[!h]
\centering
\includegraphics[scale=0.92]{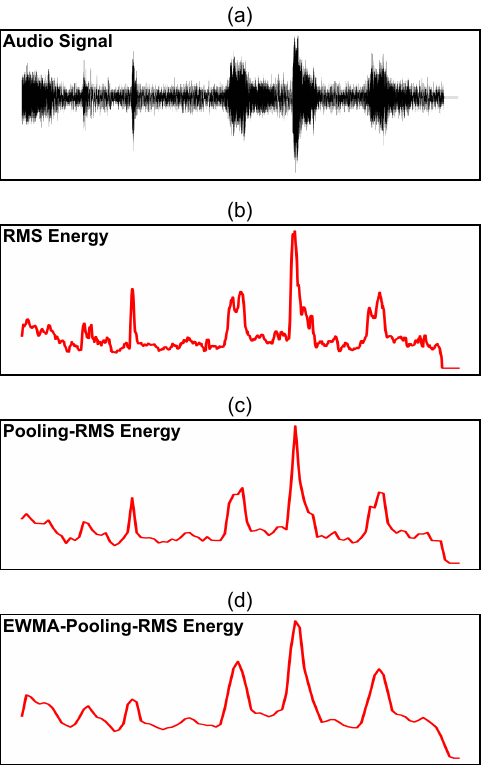} 
\caption{ 
Visualization on Loudness Signal Production Process. (a) represents the original signal and (b),(c),(d) present the signals after processed by RMS energy, APP, EWMA, respectively.
\label{fig:figure_TLM_visual}}
\end{figure}

\section{Details of Time-Loudness Module~(TLM)}

Considering the lack of smooth signals corresponding to the video-audio pairs, we propose a novel signal transformation mechanism to transform the input audio into human-like hand-drawn signals. 
As illustrated in Fig.~\ref{fig:figure_TLM_visual}, the process begins with an audio signal, as depicted in Fig.~\ref{fig:figure_TLM_visual}~(a).
Recognizing that Root Mean Square~(RMS)~\cite{panagiotakis2005speechrms} reflects the average loudness of the audio, which is helpful to generate the smoother loudness signal.
To transform the original audio signal into human-like, hand-drawn loudness signals that
forms part of the input instructions to our model, we first calculate the RMS energy as shown in Fig.~\ref{fig:figure_TLM_visual}~(b).
Compared with the original audio, RMS energy  more intuitively reflects variations in audio loudness.
Nevertheless, there are still numerous spikes in the RMS energy that is difficult to manual design.
To address these peaks, we apply Adaptive Average Pooling (AAP) to compress the signal into a lower dimension, as seen in Fig.~\ref{fig:figure_TLM_visual}~(c).
Finally, to further smooth the signal and achieve a hand-drawn quality, we employ the  Exponentially Weighted Moving Average~(EWMA)~\cite{lucas1990exponentially} to generate a signal that closely mimics a manually drawn signal, illustrated in Fig.~\ref{fig:figure_TLM_visual}~(d). 
Through these three stages of processing, the audio signal increasingly approximates a manually crafted signal.

\section{More Implementation Details}
This section provides additional clarification for the experimental details in \textit{Draw an Audio}, as presented in Tab.~\ref{tab:supp_settings}. 
It is important to emphasize that the mixed datasets are generated by combining the AudioCaps~\cite{kim2019audiocaps} and VGGSound-Caption datasets.
As the dataset size increases, we progressively increase the number of training iterations from 60k to 200k for enhancing the training process.
For a fair comparison, we following AudioLDM2~\cite{liu2023audioldm2} with the same learning rate, weight decay and optimizer.
Moreover, the term "batch size" refers specifically to the batch size per GPU with total batch size $72 \times 4$.
And we also report the approximate training time, indicating that the training cost of our model remains acceptable.

\begin{table*}[t]
    \centering
    \caption{Detailed settings. 
    This table provides a detailed overview of the specific settings and data statistics used for each dataset.}
    \vspace{0.2cm}
    \setlength{\tabcolsep}{8pt} % adjust column spacing
    % \renewcommand{\arraystretch}{0.9} % adjust row spacing
    % \begin{adjustbox}{max width=0.8\linewidth}
    \begin{tabular}{lccc} 
        \toprule 
        \textbf{Settings} & \textbf{AudioCaps}  & \textbf{VGGSound-Caption} & \textbf{Mixed Datasets} \\
        \midrule
         Number of Data & 44919 & 178730 & 223649 \\
         Duration Per Data & 10s & 10s & 10s \\
         Batch size & 72 & 72 & 72 \\
         Optimizer & AdamW & AdamW & AdamW \\
         Learning rate & 0.0001 & 0.0001 & 0.0001 \\
         Weight decay & 0.01 & 0.01 & 0.01 \\
         Iterations & 60k & 160k & 200k \\
         Training Period & $\sim$24 hours & $\sim$72 hours & $\sim$96 hours \\
         GPU Configuration & \multicolumn{3}{c}{NVIDIA A100 80G, 4 GPUs for all tasks} \\
        \bottomrule 
    \end{tabular}
    \vspace{0.2cm}
    \label{tab:supp_settings}
\end{table*}
% \emph{None} implies the complete utilization of learnable queries. 

\begin{figure*}[t]
\centering
\includegraphics[width=0.92\textwidth]{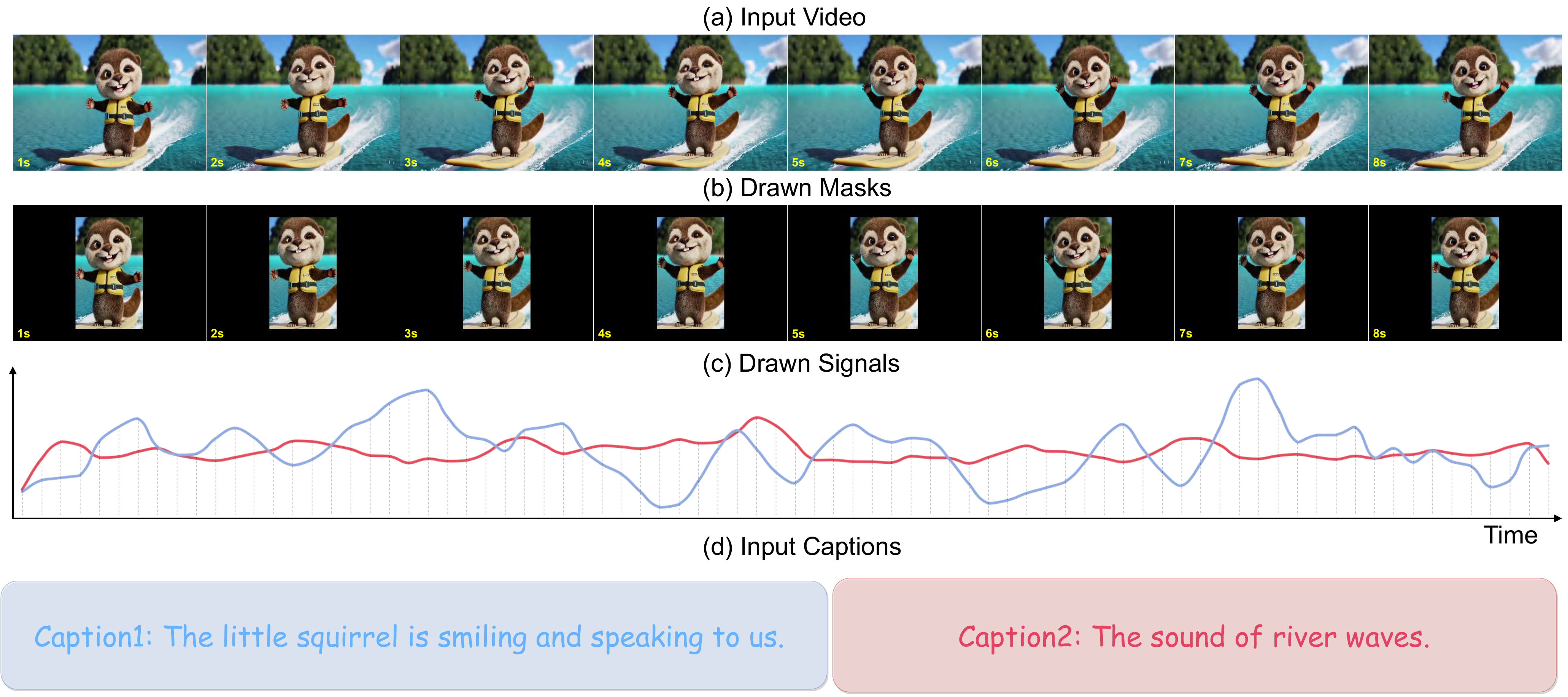} 
\caption{ 
Visualization of multiple instructions during the mixed audio production process: (a) represents the original input videos; (b) shows the masks drawn specifically for Caption 1; (c) depicts a manually designed loudness signal; and (d) corresponds to the input captions, respectively.
\label{fig:figure_Mixaudio_input}}
\end{figure*}

\begin{figure*}[ht]
\centering
\includegraphics[width=0.85\textwidth]{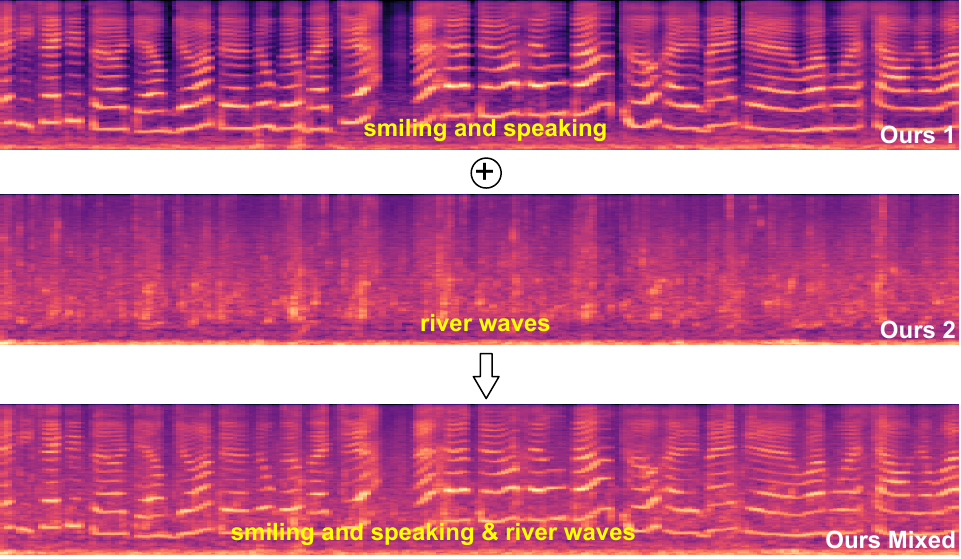} 
\caption{ 
Visualization of the mixed audio. We separately produce the sound of smiling and speaking, as well as the sound of river waves. These are then combined to create the final mixed audio.}
\label{fig:figure_Mixaudio_output}
\end{figure*}

\section{More Generation Results}

In this section, we first detail the process of the mixed audio generation comprehensively. 
Second, we present additional visualizations comparing the different methods.
Finally, we verify the generalization of our model using external videos.

\noindent \textbf{Mixed Audio Generation.} As mentioned in Fig.~1, \textit{Draw an Audio} is able to generate mixed audio in multi-stages. As illustrated in Fig~\ref{fig:figure_Mixaudio_input}, we selected the silent video with cartoon style in Sora~\cite{Brooks2024Sora} for mixed audio generation.
Specifically, we generated the sound of waves and the sound of a smiling greeting using corresponding instructions respectively. Then, as shown in Fig.~\ref{fig:figure_Mixaudio_output}, we add the two generated audios together to generate the final mixed audio.
For a better experience, the mixed audio file is available in the project page.

\begin{figure*}[ht]
\centering
\includegraphics[width=0.85\textwidth]{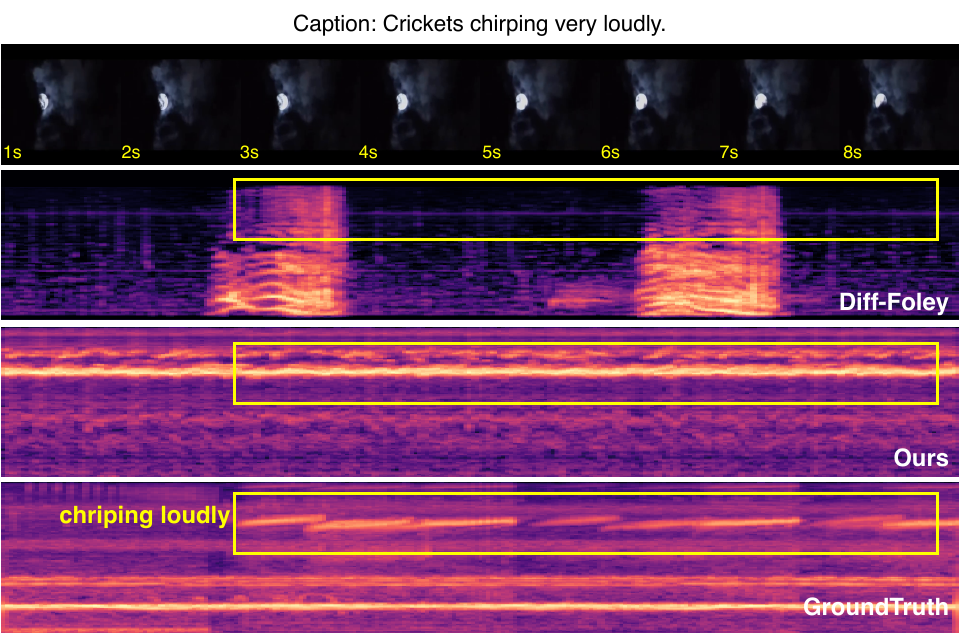}
\caption{ 
Visualization of test samples on AudioCaps dataset.}
\label{fig:supp_visual1}
\end{figure*}

\begin{figure*}[ht]
\centering
\includegraphics[width=0.85\textwidth]{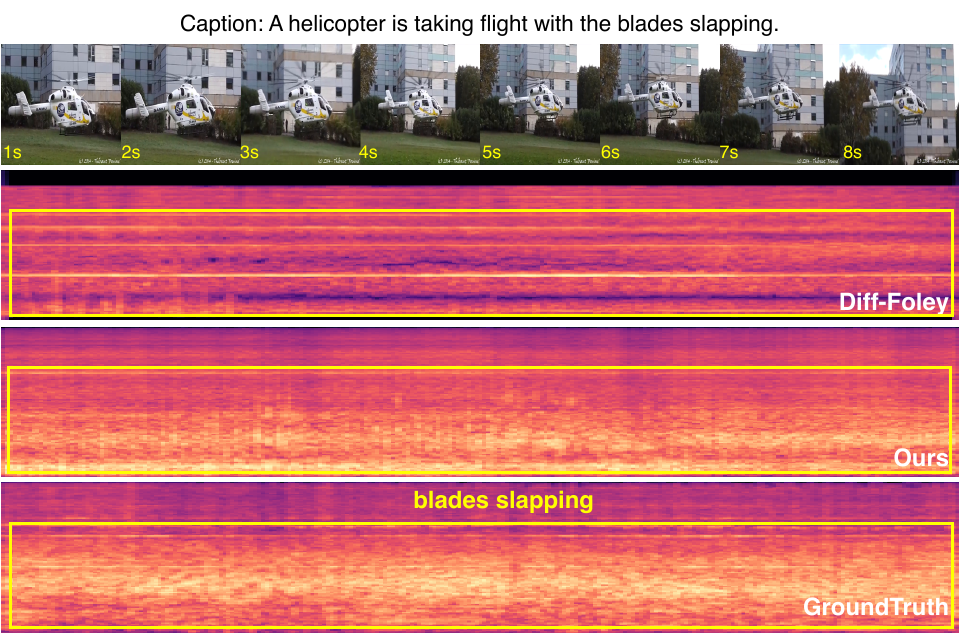}
\caption{ 
Visualization of test samples on AudioCaps dataset.}
\label{fig:supp_visual2}
\end{figure*}

\begin{figure*}[ht]
\centering
\includegraphics[width=0.85\textwidth]{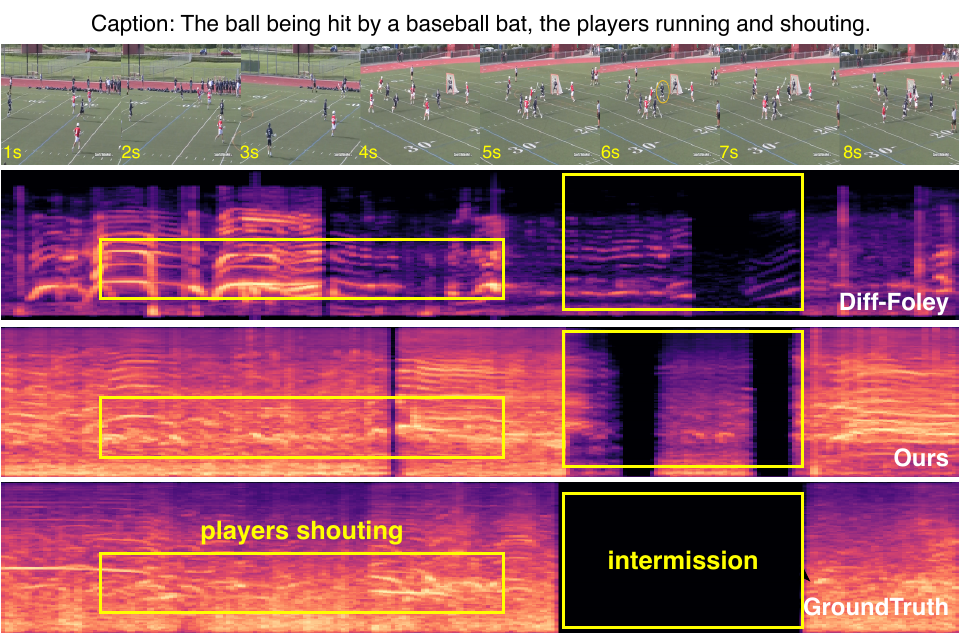}
\caption{ 
Visualization of test samples on VGGSound-Caption dataset.}
\label{fig:supp_visual3}
\end{figure*}

\begin{figure*}[ht]
\centering
\includegraphics[width=0.85\textwidth]{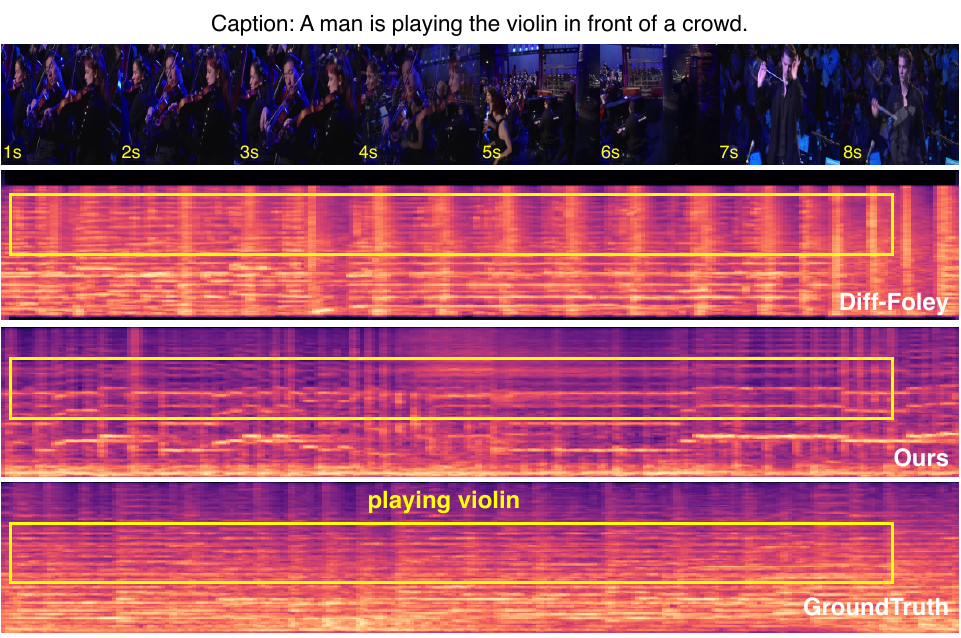}
\caption{ 
Visualization of test samples on VGGSound-Caption dataset.}
\label{fig:supp_visual4}
\end{figure*}

\noindent \textbf{More Compared Visualizations.}
We present additional comparative visualization results between our generated audios and those produced by Diff-Foley~\cite{luo2024difffoley}. 
We conduct comparisons using the AudioCaps test dataset, as illustrated in Fig.~\ref{fig:supp_visual1} and Fig.~\ref{fig:supp_visual2}, as well as the VGGSound-Caption test dataset in in Fig.~\ref{fig:supp_visual3} and Fig.~\ref{fig:supp_visual4}.
Our analysis reveals that, in comparison to Diff-Foley, the mel-spectrograms of the audio generated by our \textit{Draw an Audio} exhibit greater consistency with the original audios.
Additionally, to facilitate better comparisons, we have included more visualizations and incorporated results from SpecVQGAN~\cite{iashin2021taming} in the project page.

\noindent \textbf{External Video Visualizations.}
To further validate the generalizability of our method, we provide additional generated audio samples from external videos, including movies, cartoons, Sora~\cite{Brooks2024Sora} videos, and various daily life videos. These samples can be found in the project page.

\end{document}